\documentclass[a4paper,journal]{IEEEtran}
\usepackage[utf8]{inputenc}
\usepackage{graphicx}
\usepackage{float}
\usepackage{color, colortbl}
\usepackage{array}
\usepackage{multirow}
\usepackage{footnote}
\usepackage{cite}

\usepackage{fancyvrb}
\DefineVerbatimEnvironment{code}{Verbatim}{fontsize=\small}
\DefineVerbatimEnvironment{example}{Verbatim}{fontsize=\small}

\title{TCP/IP communication between two USRP-E110}

\author{
     \IEEEauthorblockN{Luis Sanabria-Russo\\
     \IEEEauthorblockA{Technical Report
     \\luis.sanabria@upf.edu}}
 }

\begin{document}

\maketitle

\begin{abstract}
This short report intends to provide an overview of the procedure and statistics of establishing a TCP/IP link between two USRP-E110. The testings are performed using an example GNURadio code and the networking protocol stack provided by the Linux operating system embedded in the USRP-E110.
\end{abstract}

\section{Introduction}
  During the past four months the group has been dedicating efforts towards the understanding and how to implement some of the features included in GNURadio~\cite{GNURadio}. Our testings resulted in an effective observation of the TV UHF spectrum that allowed the implementation of a TV White Spaces (TVWS) detector using energy detection technique~\cite{spec_sensing, energyDetection}.

Our current challenge relates to transmission over these detected TVWS. Fortunately, GNURadio provides a set of example code that can me used for testings and modified to achieve desired tasks.
  
\section{GNURadio code}
  The example code used to achieve network-like communication through the radio link established by the USRP-E110s is called \verb|tunnel.py|. It is usually located at \emph{/gnuradio-examples/digital/narrowband} of the GNURadio root directory.

\verb|tunnel.py| builds a TAP interface~\cite{TAP} in the kernel for tunneling Ethernet frames and uses the Operating System (OS) network protocol stack.

\begin{figure}[htbp]
  \centering
  \includegraphics[width=8cm]{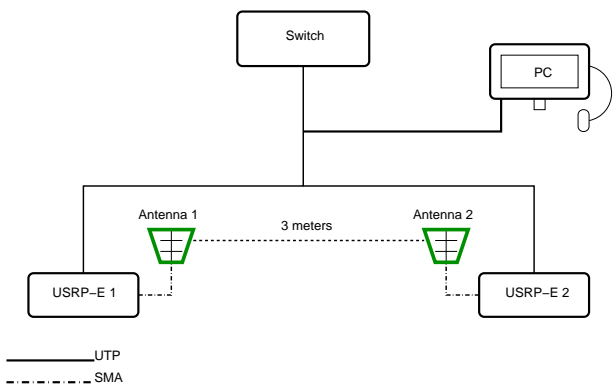}
  \caption{Testings setup.}
  \label{setup}
\end{figure}

This TAP is associated with a virtual interface that is used to test IP connectivity. Figure~\ref{setup} provides an overview of the setup used for the testings.

The execution of the code requires access to both USRP-E110s (either via USB console port or SSH). Once inside the following parameters are passed to \verb|tunnel.py|:

\begin{code}
 ./tunnel.py -f 560e6 -r 0.1e6 
 --rx-gain=20 --tx-gain=20 -m gmsk -p 2 
 --omega-relative-limit=0.00 5 
 --gain-mu=0.175 --mu=0.5 
\end{code}

, where $560$~MHz, $100$~kbps, $20$~dB, $20$~dB represent the center frequency, rate, receiver and transmitter gains accordingly. The rest of the parameters are related to PHY characteristics~\cite{PHY_parameters}.

Then, at each USRP-E110 a virtual interface (usually \verb|gr0|) should be configured with an IP address inside the same subnet by issuing the command: 

\begin{code}
  sudo ifconfig gr0 <ip address>
\end{code}

Conventional IP connectivity should be established between the two devices. 

\section{Results and future work}
  With the current approach results show around $13\%$ packet loss, delays around $60$~ms at a throughput of $100$~kbps. Also, it is possible to perform higher-layer communication with \verb|ssh| logins and remote management of the devices.

Being able to replicate this exercise allows for experimentation with medium access technologies and radio links. Future testings are pending in order to gather more accurate statistics and reduce packet loss.

\bibliographystyle{IEEEtran}
\bibliography{IEEEabrv,ref}
\end{document}